\documentclass[prb,aps,twocolumn,showpacs,nobibnotes,epsf]{revtex4}

\usepackage{graphicx}
\usepackage{dcolumn}
\usepackage{bm}
\usepackage{SIunits}

\begin{document}
\title{Superconductivity in Yb$_x$($Me$)$_y$HfNCl ($Me$ = NH$_3$ and THF)}
\author{Guojun Ye, Jianjun Ying, Yajun Yan, Xigang Luo, Peng Cheng, Ziji Xiang, Aifeng
Wang,
 Xianhui Chen} \altaffiliation{Corresponding author}
\email{chenxh@ustc.edu.cn} \affiliation{Hefei National Laboratory
for Physical Science at Microscale and Department of Physics,
University of Science and Technology of China, Hefei, Anhui 230026,
People's Republic of China\\}

\begin{abstract}
The intercalated layered nitride $\beta$-HfNCl has attracted much
attention due to the high superconducting transition temperature up
to 25.5 K. Electrons can be introduced into $\beta$-$M$NCl ($M$=Zr
and Hf) through alkali-metals intercalation to realize the
superconductivity. Here, we report the observation of
superconductivity in rare-earth metals cointercalated compounds
Yb$_x$($Me$)$_y$HfNCl with $Me$ = NH$_3$ and tetrahydrofuran (THF),
which were synthesized by the liquid ammonia method at room
temperature. The superconducting transition temperature is about 23
K and 24.6 K for Yb$_{0.2}$(NH$_3$)$_y$HfNCl and
Yb$_{0.3}$(NH$_3$)$_y$HfNCl, respectively. Replacing the NH$_3$ with
a larger molecule THF, superconducting transition temperature
increases to 25.2 K in Yb$_{0.2}$(THF)$_y$HfNCl, which is almost the
same as the highest $T_{\rm c}$ reported in the alkali-metals
intercalated HfNCl superconductors. The $T_{\rm c}$ of
Yb$_{0.2}$(THF)$_y$HfNCl is apparently suppressed by pressure up to
0.5 GPa, while the pressure effect on $T_{\rm c}$ becomes very small
above 0.5 GPa. The liquid ammonia method is proved to be an
effective synthetic method to intercalate metal ions into HfNCl. Our
results suggest that the superconductivity in these layered
intercalated superconductors nearly does not rely on the
intercalated metal ions, even magnetic ion.
\end{abstract}
\maketitle

High-$T_{\rm c}$ superconductivity has been observed in layered
cuprates and recently discovered iron-based
superconductors.\cite{Bednorz,Hosono,ChenXH} The proximity to
magnetically ordered states for these systems suggested that the
magnetic interactions are the crucial force for Cooper pairing in
such high-$T_{\rm c}$ superconductivity, which gives rise to the
unconventional nature of the superconductivity in these systems.
Superconductivity in other various exotic materials, such as
Na$_x$CoO$_2$, Sr$_2$RuO$_4$ and heavy-fermion systems, was also
found to be intimate to magnetism.\cite{Foo,Mackenzie,5} However,
for another type of superconducting material of the layered
metallonitride halides ($M$N$X$, $M$ = Ti, Zr, Hf; $X$ = Cl, Br, I)
with the maximum $T_{\rm c}$ as high as 25.5 K,\cite{Yamanaka1}
their parent compounds are band insulators and superconductivity
seems to have no correlation with magnetism.\cite{Kuroki} There
exist two types of the layered nitride compounds: one is the
(FeOCl)-type structure (so called $\alpha$ structure) with the 2D
metal-nitrogen ($M$N) layer of rectangular lattice; the other is the
(SmSI)-type one (so called $\beta$ structure) with the 2D $M$N layer
of honeycomb lattice.\cite{Sugimoto} For the former, K$_x$TiNCl was
reported to display superconductivity with $T_{\rm c}$ = 16
K.\cite{KTiNCl} For the latter, usually with $M$ = Hf, Zr and $X$ =
Cl, maximum of $T_{\rm c}$ = 25.5 K has been achieved in
Li$_x$(THF)$_y$HfNCl.\cite{Yamanaka1} The parent compounds of the
latter, so-called $\beta$-$M$NCl, consist of alternative stacking of
honeycomb MN bilayer sandwiched by Cl bilayer.\cite{Juza}
Superconductivity is usually induced through doping charge carriers
by means of alkali-metal intercalation or producing the $Cl$
deficiency.\cite{Yamanaka2,Zhu} Unlike the large pressure effect on
$T_{\rm c}$ observed in cuprates or iron-based superconductors, the
$T_{\rm c}$ in this type of superconductors decreases slightly as
the pressure increases\cite{Taguchi3, Shamotoa}. However, for
cointercalated $\beta$-ZrNCl and $\beta$-HfNCl, the interlayer
spacing would strongly affects the superconducting transition
temperature. Increase of the basal spacing would lead to the
reduction of the tiny warping along the K$_z$ direction, thus to
increase the nesting of the Fermi surface\cite{Takano}. It is
assumed that the modification of the Fermi surface would increase
the pairing interaction among the electrons, which would enhance
$T_{\rm c}$ and the maximum $T_{\rm c}$ is found when the basal
spacing increased to approximate 15 ${\rm \AA}$ in this type of
materials\cite{Yamanaka}.

In two dimensional superconductors, the spin fluctuation may lead to
unconventional pairing and high-$T_{\rm c}$ superconductivity might
emerge\cite{Moriya, Monthoux}. Nuclear magnetic resonance (NMR)
\cite{Tou} and muon spin relaxation ($\mu$SR)
experiments\cite{Uemura,Ito} revealed the two-dimensional nature of
superconductivity in this intercalated layered nitride $M$NCl. The
$M$N bilayer honeycomb structure is thought to play a main role for
the happening of superconductivity.\cite{Kuroki} The NMR knight
shift suggested a spin-singlet pairing, \cite{Tou1} and tunneling
spectroscopy \cite{Ekino1,Ekino2} as well as specific heat
\cite{Taguchi} revealed a fully open s-wave-like gap. The
tunneling-current measurements \cite{Takasaki,Takasaki1} and
specific-heat \cite{Tou3} gave a quite large superconducting gap
with the ratio 2$\Delta$/$k_{\rm B}T$ $\approx$ 4.6-5.2 or even
larger, suggesting the strong coupling superconductivity. However,
some recent results, such as the anisotropic gap in large doping
level inferred by $\mu$SR \cite{Hiraishi} and the absence of
coherence peak in spin-lattice-relaxation rate revealed by NMR
experiment \cite{Tou2}, suggested the unconventional pairing
mechanisms. Moreover, relatively high $T_{\rm c}$ with extremely low
density of states at Fermi level\cite{Taguchi,Tou3,Weht}, weak
electron-phonon coupling \cite{Tou1,Tou3,Weht,Kitora} and small
isotope effect \cite{Tou4,Taguchi2} also favor the unconventional
pairing mechanisms in these intercalated $\beta$-$M$NCl
superconductors. The mystery of the superconductivity for the
intercalated $\beta$-$M$NCl compounds is still in debt.

In this report, we reported the discovery of superconductivity by
cointercalating magnetic rare earth ion of ytterbium with NH$_3$ or
THF molecule in HfNCl. Ytterbium was cointercalated with NH$_3$
between HfNCl layers by the liquid ammonia method at room
temperature, instead of previous methods by reacting in
alkali-organic salt/organic solution,\cite{Yamanaka1}
electrochemical intercalation,\cite{Dompablo} using solid-state
reaction with K$_3$N\cite{Zhu2}. Superconductivity with $T_{\rm c}$
of $\sim 23$ K or $\sim 24.6$ K is discovered in
Yb$_x$($NH_3$)$_y$HfNCl depending on the Yb content. THF molecule
can be also cointercalated with Ytterbium into HfNCl, and the
superconductivity with $T_{\rm c}$ as high as 25.2 K was observed
from magnetic susceptibility, which is nearly the same as the
reported maximum $T_{\rm c}$ in the alkali-metal intercalated HfNCl
compounds. The pressure effect of this sample is negative, d$T_{\rm
c}$/dP is about -0.6 GPa/K below 0.5 GPa and it becomes -0.16 GPa/K
above 0.5 GPa.

\begin{figure}
\begin{center}
\centerline{\includegraphics[width=.5\textwidth]{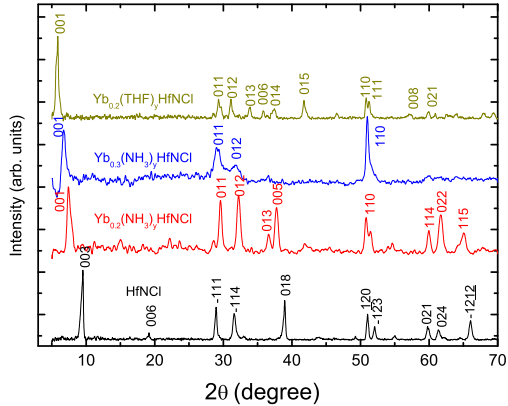}}
\caption{The x-ray diffraction patterns of pristine $\beta$-HfNCl
and the superconducting samples of Yb$_{0.2}$(NH$_3$)$_y$HfNCl,
Yb$_{0.3}$(NH$_3$)$_y$HfNCl and Yb$_{0.2}$(THF)$_y$HfNCl,
respectviely.}\label{fig1}
\end{center}
\end{figure}

\begin{figure}
\begin{center}
\centerline{\includegraphics[width=.5\textwidth]{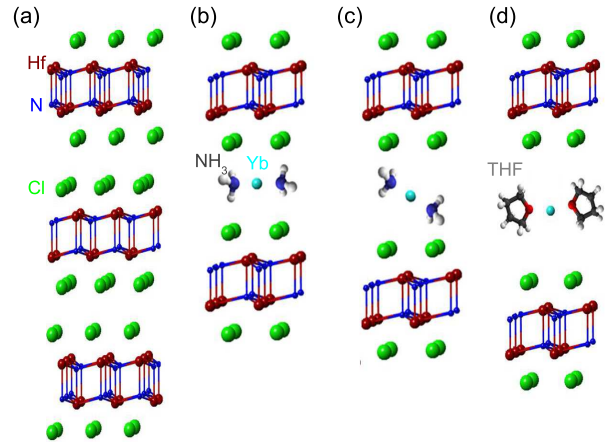}}
\caption{The schematic structural models for (a): the pristine
HfNCl; (b): $Yb_{0.2}(NH_3)_y$HfNCl; (c): $Yb_{0.3}(NH_3)_y$HfNCl
 and (d): $Yb_{0.2}(THF)_y$HfNCl, respectively.}\label{fig2}
\end{center}
\end{figure}

\begin{table*}[t]
\caption{Lattice parameters and $T_{\rm c}$ of Yb$_x$($Me$)$_y$HfNCl
($Me$ = NH$_3$ and THF).}
\begin{center}
\begin{tabular}{{@{\vrule height 10.5pt depth4pt  width0pt}lrcccc}}
       &$\beta$-HfNCl & Yb$_{0.2}$(NH$_3$)$_y$HfNCl & Yb$_{0.3}$(NH$_3$)$_y$HfNCl & Yb$_{0.2}$(THF)$_y$HfNCl &\\   \hline
      space group & R$\overline{3}$m & P$\overline{3}$m & P$\overline{3}$m & P$\overline{3}$m \\
      a (${\rm \AA}$) & 3.58 & 3.59 & 3.59 & 3.59 \\
      c (${\rm \AA}$) & 27.71 & 11.95 & 13.20 & 15.05 \\
      d-spacing (${\rm \AA}$) & 9.24 & 11.95 & 13.20 & 15.05 \\
      $T_{\rm c}$ (K) & & 23 & 24.6 & 25.2 \\
 \hline
\end{tabular}
\end{center}
\end{table*}

\begin{figure}
\centering
\centerline{\includegraphics[width=.5\textwidth]{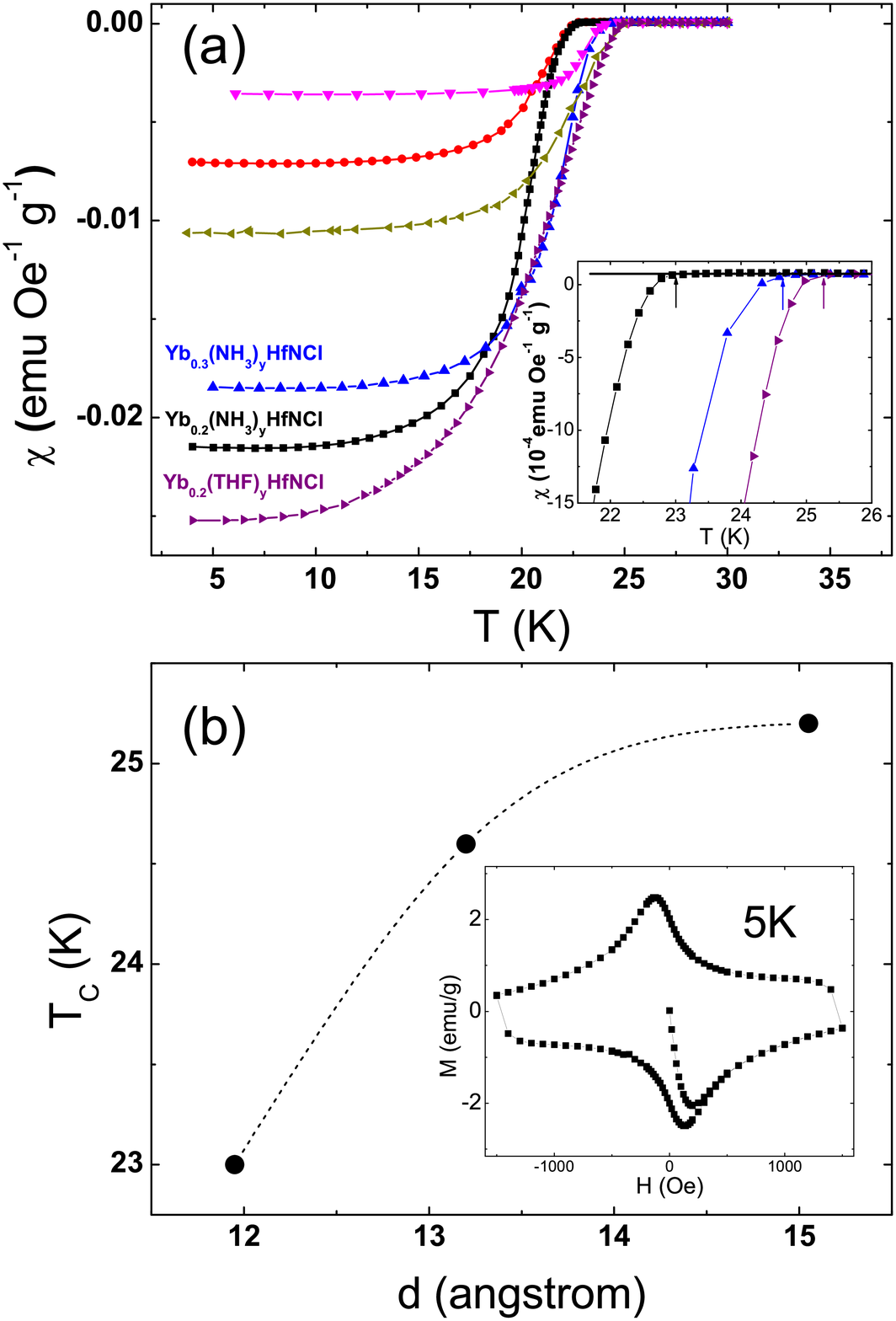}}
\caption{(a): The ZFC and FC susceptibility taken at 10 Oe for
Yb$_{0.2}$(NH$_3$)$_y$HfNCl, Yb$_{0.3}$(NH$_3$)$_y$HfNCl and
Yb$_{0.2}$(THF)$_y$HfNCl. The inset shows the enlarged area around
$T_{\rm c}$. (b): Interlayer spacing d dependence of $T_{\rm c}$ for
all the superconducting samples. The inset shows the isothermal
magnetization hysteresis of Yb$_{0.2}$(NH$_3$)$_y$HfNCl taken at
5K.}\label{fig3}
\end{figure}

\begin{figure}
\centerline{\includegraphics[width=.5\textwidth]{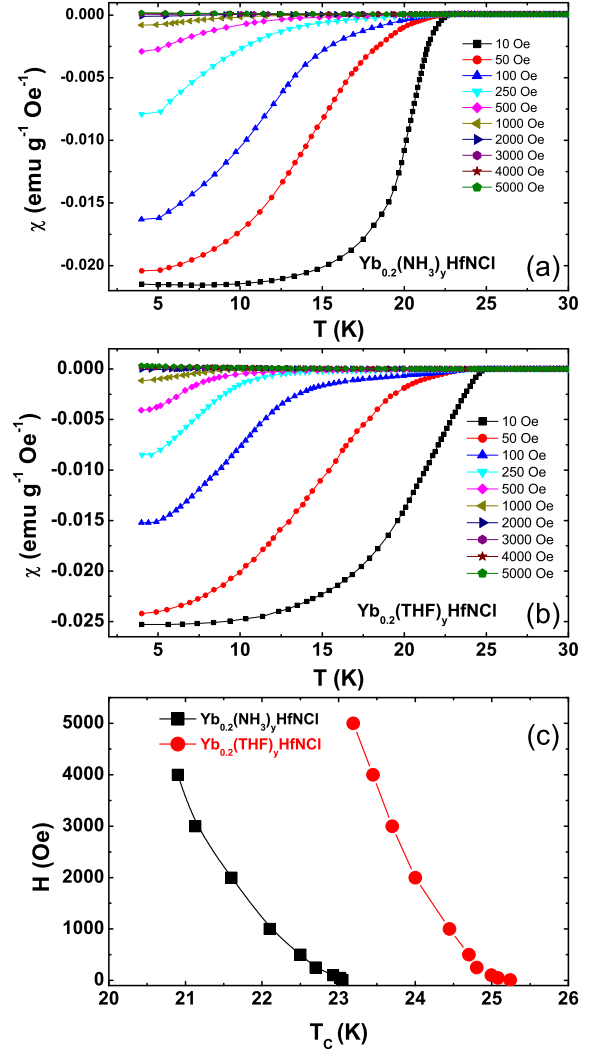}}
\caption{Temperature dependence of susceptibility for the
superconducting samples of (a): Yb$_{0.2}$(NH$_3$)$_y$HfNCl and (b):
Yb$_{0.2}$(THF)$_y$HfNCl in the ZFC measurements under different
magnetic fields. (c): The $H_{\rm c2}$ versus $T_{\rm c}$ for the
samples of Yb$_{0.2}$(NH$_3$)$_y$HfNCl and
Yb$_{0.2}$(THF)$_y$HfNCl.}\label{fig4}
\end{figure}

\begin{figure}
\centerline{\includegraphics[width=.5\textwidth]{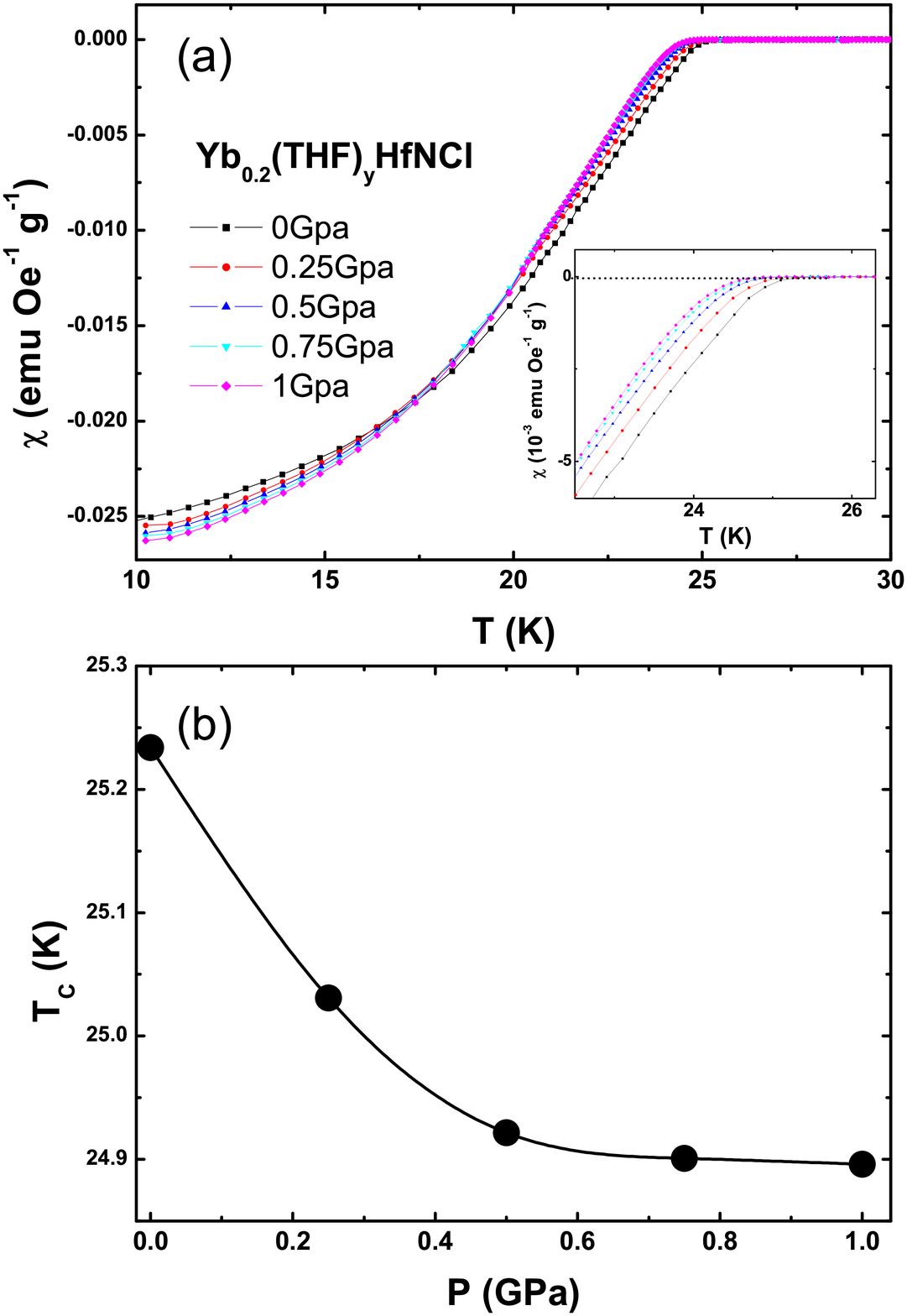}}
\caption{(a): Temperature dependence of susceptibility for the
sample Yb$_{0.2}$(THF)$_y$HfNCl in the ZFC measurements under
various pressures. The inset is the enlarged area around $T_{\rm
c}$. (b): Pressure dependence of $T_{\rm c}$ for the sample
Yb$_{0.2}$(THF)$_y$HfNCl.}\label{fig5}
\end{figure}
Figure 1 shows the X-ray diffraction (XRD) patterns of
$\beta$-HfNCl, Yb$_{0.2}$(NH$_3$)$_y$HfNCl,
Yb$_{0.3}$(NH$_3$)$_y$HfNCl and Yb$_{0.2}$(THF)$_y$HfNCl using Cu
$K_\alpha$ radiations. The XRD patten of pristine $\beta$-HfNCl can
be well indexed based on the space group R$\overline{3}$m, and the
lattice parameters are determined to be a=3.58 ${\rm \AA}$ and
c=27.71 ${\rm \AA}$, being consistent with the previous
report\cite{Yamanaka1}. In comparison with that of the pristine
$\beta$-HfNCl, the XRD patterns of Yb$_{0.2}$(NH$_3$)$_y$HfNCl,
Yb$_{0.3}$(NH$_3$)$_y$HfNCl and Yb$_{0.2}$(THF)$_y$HfNCl can be
indexed based on the space group P$\overline{3}$m. The lattice
parameters are determined to be a=3.59 ${\rm \AA}$ and c=11.95 ${\rm
\AA}$ for Yb$_{0.2}$(NH$_3$)$_y$HfNCl, and a=3.59 ${\rm \AA}$ and
c=13.20 ${\rm \AA}$ for Yb$_{0.3}$(NH$_3$)$_y$HfNCl, respectively.
The lattice parameters change to be a=3.59 ${\rm \AA}$ and c=15.05
${\rm \AA}$ for Yb$_{0.2}$(THF)$_y$HfNCl. The lattice parameters in
the ab plane are almost unchanged for all the intercalated samples,
but the stacking pattern of the layers is changed so much, leading
to the change in the space group from R$\overline{3}$m to
P$\overline{3}$m. The d-spacing between HfNCl layers increases from
9.24 ${\rm \AA}$ for pristine $\beta$-HfNCl to 11.95${\rm \AA}$ and
13.20${\rm \AA}$ for the superconducting Yb$_{0.2}$(NH$_3$)$_y$HfNCl
and Yb$_{0.3}$(NH$_3$)$_y$HfNCl, and to 15.05${\rm \AA}$ for
superconducting Yb$_{0.2}$(THF)$_y$HfNCl, respectively. The
interlayer spacing $d$ between $M$NCl ($M$=Zr or Hf) layers is
strongly dependent on the amounts and types of metal ions, and the
cointercalated solvent molecules in intercalated $M$NCl
superconductors. The c-axis lattice parameter of c=11.95 ${\rm \AA}$
for Yb$_{0.2}$(NH$_3$)$_y$HfNCl is nearly the same as c=12.1 ${\rm
\AA}$ of Li$_{0.37}$(NH$_3$)$_y$HfNCl\cite{Takano}. It indicates
that the stacking structure for NH$_3$ cointercalated HfNCl with Yb
should be similar to that of Li$_{0.37}$(NH$_3$)$_y$HfNCl. While for
Yb$_{0.3}$(NH$_3$)$_y$HfNCl, the d-spacing increases to 13.20${\rm
\AA}$, which is possibly due to the different Yb amount and
orientation of NH$_3$. The spacing between $M$NCl ($M$=Zr or Hf)
layers increases to 14.9 or 18.5 ${\rm \AA}$ for
Li$_x$(THF)$_y$ZrNCl,\cite{Yamanaka3} while to 13.6 or 18.7${\rm
\AA}$\cite{Takano} for Li$_x$(THF)$_y$HfNCl. The different spacing
between MNCl layers strongly depends on the amounts of lithium and
orientation of THF. The c-axis parameter increases to 15.05 ${\rm
\AA}$ for the Yb$_{0.2}$(THF)$_y$HfNCl, which is very close to 14.9
${\rm \AA}$ for Li$_x$(THF)$_y$ZrNCl. It suggests that the stacking
structure of the Yb$_{0.2}$(THF)$_y$HfNCl is the same as that the
Li$_x$(THF)$_y$ZrNCl. The Schematic structural models for the
pristine HfNCl, Yb$_{0.2}$(NH$_3$)$_y$HfNCl,
Yb$_{0.3}$(NH$_3$)$_y$HfNCl and Yb$_{0.2}$(THF)$_y$HfNCl are
proposed as shown in Fig.2, respectively.

Temperature dependence of zero field cooling (ZFC) and field cooling
(FC) magnetic susceptibilities for the superconducting
Yb$_{0.2}$(NH$_3$)$_y$HfNCl, Yb$_{0.3}$(NH$_3$)$_y$HfNCl and
Yb$_{0.2}$(THF)$_y$HfNCl are shown in Fig.3 (a). The zero field
cooling (ZFC) susceptibilities shown in the inset of Fig. 3(a)
indicate a clear superconducting transition at about 23 K for
Yb$_{0.2}$(NH$_3$)$_y$HfNCl, at 24.6 K for
Yb$_{0.3}$(NH$_3$)$_y$HfNCl and at 25.2 K for
Yb$_{0.2}$(THF)$_y$HfNCl, respectively. Interlayer spacing d
dependence of $T_{\rm c}$ for all the superconducting samples is
shown in Fig. 3(b), $T_{\rm c}$ slightly increases from 23 K to 25.2
K as interlayer spacing $d$ increases from 11.95 ${\rm \AA}$ to
15.05 ${\rm \AA}$. A similar behavior has been observed in
Li$_x$$Me_y$HfNCl($Me$=NH$_3$ and THF)\cite{Takano}. The inset of
Fig.3(b) shows the isothermal magnetization hysteresis for
Yb$_{0.2}$(NH$_3$)$_y$HfNCl at 5 K. Similar behavior in the M-H is
observed for the samples of Yb$_{0.3}$(NH$_3$)$_y$HfNCl and
Yb$_{0.2}$(THF)$_y$HfNCl. The lower critical field (H$_{c1}$) for
all the superconducting samples are around 80 Oe, which are the same
as that of alkali-metal cointercalated HfNCl\cite{Yamanaka3}.
Lattice parameters and $T_{\rm c}$ of Yb$_x$($Me$)$_y$HfNCl ($Me$ =
NH$_3$ and THF) are summarized in Table 1.

Figure 4(a) and (b) show the temperature dependence of the
susceptibility in ZFC measurements under various magnetic fields for
Yb$_{0.2}$(NH$_3$)$_y$HfNCl and Yb$_{0.2}$(THF)$_y$HfNCl,
respectively. $T_c$ and the diamagnetic signal are gradually
suppressed, and the superconducting transition becomes significantly
broad with the application of magnetic fields. Within the
weak-coupling BCS theory, the upper critical field $H_{\rm c2}$ at
$T$=0 K can be determined by the Werthamer-Helfand-Hohenberg (WHH)
equation\cite{Werthamer} $H_{\rm c2}(0)=0.693[-(dH_{\rm
c2}/dT)]_{T_{\rm c}}T_{\rm c}$. Using the data of $H_{\rm c2}$(T)
derived from the susceptibility measurement, one obtains
$[-(dH_{c2}/dT)]_{T_{\rm c}}$ to be about 0.25 T/K and 0.38 T/K for
Yb$_{0.2}$(NH$_3$)$_y$HfNCl and Yb$_{0.2}$(THF)$_y$HfNCl,
respectively. Thus, the $H_{\rm c2}(0)$ can be estimated to be 4 T
and 6.6 T for Yb$_{0.2}$(NH$_3$)$_y$HfNCl and
Yb$_{0.2}$(THF)$_y$HfNCl, respectively.

Figure 5(a) shows the temperature dependence of the susceptibility
in ZFC measurements for Yb$_{0.2}$(THF)$_y$HfNCl under various
pressures. The inset of Fig. 5(a) shows the enlarged area around
$T_{\rm c}$. The $T_{\rm c}$ is defined as the temperature at which
the susceptibility starts to decrease. The pressure dependence of
$T_{\rm c}$ was shown in Fig. 5(b). $T_{\rm c}$ decreases with
increasing the pressure. $T_{\rm c}$ deceases at a relative quick
speed with d$T_{\rm c}$/dP=-0.6 GPa/K below 0.5 GPa. While the
pressure effect becomes very small above 0.5 GPa with d$T_{\rm
c}$/dP=-0.16 GPa/K. Such behavior is similar to the observation in
the alkali-metals intercalated HfNCl and ZrNCl\cite{Taguchi3,
Shamotoa}.

Electron-doping of $\beta$-HfNCl is usually realized by the
intercalation of alkali metals or cointercalation of alkali metals
with molecules. Here, we report the superconductivity in
electron-doped HfNCl by cointercalation of rare-earth magnetic ion
with molecules. It is striking that the maximum $T_{\rm c}$ of 25.2
K observed in Yb$_{0.2}$(THF)$_y$HfNCl is almost the same as the
highest $T_c$ in the alkali metals cointercalation with THF. It
indicates that superconductivity in the intercalated HfNCl does not
rely on the different intercalated ions, even magnetic ion. It is
intriguing that the intercalation of magnetic ion of rare-earth
metal Yb does not affect the superconductivity relative to the
intercalation of alkali-metal ion. It indicates that magnetism does
not suppress the superconductivity, being an evidence for
unconventional superconductivity. The $T_{\rm c}$ increases from 23
K to 25.2 K with increasing the interlayer spacing from 11.95 ${\rm
\AA}$ for Yb$_{0.2}(NH_3)_y$HfNCl to 15.05 ${\rm \AA}$ for
Yb$_{0.2}$(THF)$_y$HfNCl. Such slight enhancement of $T_c$ induced
by large increase of the interlayer spacing indicates the good
two-dimensional electronic system for the intercalated HfNCl
superconductors, which could be the reason why superconductivity in
the intercalated HfNCl does not rely on the intercalated ions. The
liquid ammonia method is proved to be a good way to intercalate
metal ions into MNCl (M=Zr and Hf) to introduce electron in the
system and realize superconductivity. An interesting question is
whether $T_c$ could be raised to above 25.5 K in the intercalated
HfNCl system or not.

\textbf{\large Materials and Methods} $\beta$-HfNCl was synthesized
by reacting of Hf powder and gasified NH$_4$Cl in the environment of
ammonia at 923 K for 30 minutes, then the product was sealed in a
quartz tube followed by a vapor transport recrystallized process
from low temperature side to high temperature side at the
temperature gradient of 1023 K to 1123 K with the aid of a small
amount of NH$_4$Cl as transport agent. We can obtained two types of
Yb$_x$(NH$_3$)$_y$HfNCl by adjusting the Yb content. 0.1 gram of
recrystallized HfNCl together with 0.053 or 0.068 gram of ytterbium,
then the mixture were loaded in a 50 ml autoclave which was cooled
by liquid nitrogen, the autoclave was slowly filled with 15 ml
liquid ammonia and sealed. The sealed autoclave was kept at room
temperature for 1-3 days before it was opened and dried in the glove
box. The products were rinsed by using liquid ammonia to eliminate
soluble impurities, thus we can obtain the final product. The actual
Yb concentration (x) of these two samples were determined by
inductively coupled plasma atomic emission spectroscopy (ICP-AES),
and the actual x values are 0.2 and 0.3, respectively.
Yb$_{0.2}$(THF)$_y$HfNCl can be synthesized by immersing the
Yb$_{0.2}$(NH$_3$)$_y$HfNCl powder into THF solution for 1-2 days,
while we can not obtain Yb$_{0.3}$(THF)$_y$HfNCl by the same method.
All the experiments were performed under Ar atmosphere to prevent it
from air and water contamination. The x-ray diffraction (XRD) was
carried out with samples sealed in capillaries that were made of
special glass No. 10 and purchased from Hilgenberg GmbH. The
magnetization measurement was performed by using SQUID MPMS-5T
(Quantum Design). The magnetization under pressure was measured by
incorporating a copper每beryllium pressure cell (EasyLab) into SQUID
MPMS (Quantum Design). The sample was firstly placed in a teflon
cell (EasyLab) with coal oil (EasyLab) as the pressure media. Then,
the teflon cell was set in the copper-beryllium pressure cell for
magnetization measurement.


\begin{thebibliography}{}
\bibitem{Bednorz}
J. G. Bednorz, and K. A. M$\ddot{u}$ller, (1986) Possible high
$T_{\rm c}$ superconductivity in the Ba-La-Cu-O system. Z. Phys. B
\textbf{64}, 189.

\bibitem{Hosono}
Y. Kamihara, H. Hiramatsu, M. Hirano, R. Kawamura, H. Yanagi, T.
Kamiya, and H. Hosono, (2006) Iron-based layered superconductor
La(O$_{1-x}$F$_x$)FeAs (x=0.05每0.12) with $T_{\rm c}$=26 K. J. Am.
Chem. Soc. \textbf{128}, 10012.

\bibitem{ChenXH}
X. H. Chen, T. Wu, G. Wu, R. H. Liu, H. Chen, and D. F. Fang,(2008)
Superconductivity at 43 K in SmFeAsO$_{1-x}$F$_x$. Nature (London)
\textbf{453}, 761.

\bibitem{Foo}
M. L. Foo, Y. Y. Wang, S. Watauchi, H. W. Zandbergen, T. He, R. J.
Cava, and N. P. Ong, (2004) Charge Ordering, Commensurability, and
Metallicity in the Phase Diagram of the Layered Na$_x$CoO$_2$. Phys.
Rev. Lett. \textbf{92}, 247001.

\bibitem{Mackenzie}
A. P. Mackenzie, Y. Maeno, (2003) The superconductivity of
Sr$_2$RuO$_4$ and the physics of spin-triplet pairing. Rev. Mod.
Phys. \textbf{75}, 657.

\bibitem{5}
P. Coleman, {\sl Handbook of Magnetism and Advanced Magnetic
Materials}. $Edited$ by Helmut Kronmuller and Stuart Parkin. Vol 1:
Fundamentals and Theory. ($John~Wiley~and~Sons$), 95-148 (2007).

\bibitem{Yamanaka1}
S. Yamanaka, K. Hotehama, and H. Kawaji, (1998) Superconductivity at
25.5K in electron-doped layered hafniumnitride. Nature \textbf{392},
580.

\bibitem{Kuroki}
K. Kuroki, (2010) Spin-fluctuation-mediated d+id∩ pairing mechanism
in doped $\beta$-MNCl (M=Hf,Zr) superconductors. Phys. Rev. B
\textbf{81}, 104502.

\bibitem{Sugimoto}
A. Sugimoto, K. Shohara, T. Ekino, Z. F. Zheng, and S. Yamanaka,
(2012) Nanoscale electronic structure of the layered nitride
superconductors $\alpha$-K$_x$TiNCl and $\beta$-HfNCl$_y$ observed
by scanning tunneling microscopy and spectroscopy. Phys. Rev. B
\textbf{85}, 144517.

\bibitem{KTiNCl}
S.Yamanaka, T.Yasunaga,K.Yamaguchi, and M. Tagawa, (2009) Structure
and superconductivity of the intercalation compounds of TiNCl with
pyridine and alkali metals as intercalants. J. Mater. Chem.
\textbf{19}, 2573.

\bibitem{Juza}
R. Juza, and H. Friedrichsen, (1964) Die Kristallstruktur von
$\beta$-ZrNCl und $\beta$-ZrNBr. Z. Anorg. Allg. Chem. \textbf{332},
173每178.

\bibitem{Yamanaka2}
S. Yamanaka, H. Kawaji, K. Hotehama, and M. Ohashi, (1996) A new
layer-structured nitride superconductor Lithium-intercalated
$\beta$-zirconium nitride chloride, Li$_x$ZrNCl. Adv. Mater.
\textbf{9}, 771.

\bibitem{Zhu}
L. Zhu and S. Yamanaka, (2003) Preparation and Superconductivity of
Chlorine-Deintercalated Crystals $\beta$-MNCl$_{1-x}$ (M = Zr, Hf).
Chem. Mater. \textbf{15}, 1897.

\bibitem{Taguchi3}
Y. Taguchi, M. Hisakabe, Y. Ohishi, S. Yamanaka, and Y. Iwasa,
(2004) High-pressure study of layered nitride superconductors. Phys.
Rev. B \textbf{70}, 104506.

\bibitem{Shamotoa}
S. Shamotoa,  K.  Iizawaa,  T. Koiwasakib,  M. Yasukawab,  S.
Yamanakab,  O.  Petrenkoc,  S. M. Benningtonc, H. Yoshidad,  K.
Ohoyamad,  Y. Yamaguchid,  Y. Qnoa,  Y. Miyazakia  and  T.
Kajitania, (2000) Pressure  Effect  and Neutron  Scattering  Study
on A$_x$HfNCl (A; Alkali  Metals  and  Organic Molecules). Physica C
\textbf{341每348} 747每748.


\bibitem{Takano}
T. Takano, T. Kishiume, Y. Taguchi, and Y. Iwasa, (2008)
Interlayer-Spacing Dependence of $T_{\rm c}$ in Li$_x$$M_y$HfNCl (M:
Molecule) Superconductors. Phys. Rev. Lett. \textbf{100}, 247005.

\bibitem{Yamanaka}
Shoji Yamanaka, (2010) Intercalation and superconductivity in
ternary layer structured metal nitride halides (MNX: M = Ti, Zr, Hf;
X = Cl, Br, I). J. Mater. Chem. \textbf{20}, 2922每2933.

\bibitem{Moriya}
T. Moriya, Y. Takahashi and K. Ueda, (1990) Antiferromagnetic Spin
Fluctuations and Superconductivity in Two-Dimensional Metals -A
Possible Model for High $T_{\rm c}$ Oxides. J. Phys. Soc. Jpn.
\textbf{59}, 2905.

\bibitem{Monthoux}
P. Monthoux, D. Pines, (1994) Spin-fluctuation-induced
superconductivity and normal-state properties of YBa$_2$Cu$_3$O$_7$.
Phys. Rev. B \textbf{49}, 4261每4278.

\bibitem{Tou}
H. Tou, Y. Maniwa, T. Koiwasaki, and S. Yamanaka, (2000) Evidence
for quasi-two-dimensional superconductivity in electron-doped
Li$_{0.48}$(THF)$_y$HfNCl. Phys. Rev. B \text{63}, 020508(R).

\bibitem{Uemura}
Y. J. Uemura, Y. Fudamoto, I. M. Gat, M. I. Larkin, G. M. Luke, J.
Merrin, K. M. Kojima, K. Itoh, S. Yamanaka, R. H. Heffner, and D. E.
MacLaughlin, (2000) $\mu$SR studies of intercalated HfNCl
superconductor. Physica B \textbf{289-290}, 389.

\bibitem{Ito} 
T. Ito, Y. Fudamoto, A. Fukaya, I. M. Gat-Malureanu, M. I. Larkin,
P. L. Russo, A. Savici, Y. J. Uemura, K. Groves, R. Breslow, K.
Hotehama, S. Yamanaka, P. Kyriakou, M. Rovers, G. M. Luke, and K. M.
Kojima, (2004) Two-dimensional nature of superconductivity in the
intercalated layered systems Li$_x$HfNCl and Li$_x$ZrNCl: Muon spin
relaxation and magnetization measurements. Phys. Rev. B \textbf{69},
134522.

\bibitem{Tou1} 
H. Tou, S. Oshiro, H. Kotegawa, Y. Taguchi, Y. Kasahara, I. Iwasa,
(2010) $^{15}$N NMR Studies of Layered Nitride Superconductor
Li$_x$ZrNCl. Physica C\textbf{470} S658每S659.

\bibitem{Ekino1}
T. Ekino, T. Takasaki, H. Fujii, and S. Yamanaka, (2003) Tunneling
spectroscopy of the eiectron doped layered superconductor
Li$_{0.48}$(THF)$_{0.3}$HfNCl. Physica C \textbf{388-389}, 573.

\bibitem{Ekino2} 
T. Ekino, T. Takasaki, T. Muranaka, H. Fujii, J. Akimitsu, and S.
Yamanaka, Tunneling Spectroscopy of MgB$_2$ and
Li$_{0.5}$(THF)$_y$HfNCl. Physica B \textbf{328}, 23 (2003).

\bibitem{Taguchi}
Y. Taguchi, M. Hisakabe, and Y. Iwasa, (2005) Specific Heat
Measurement of the Layered Nitride Superconductor Li$_x$ZrNCl. Phys.
Rev. Lett. \textbf{94}, 217002.

\bibitem{Takasaki} 
T. Takasaki, T. Ekino, H. Fujii, and S. Yamanaka, (2005) Tunneling
spectroscopy of deintercalated layered nitride superconductors
ZrNCl$_{0.7}$. J. Phys. Soc. Jpn. \textbf{74}, 2586.

\bibitem{Takasaki1}
T. Takasaki, T. Ekino, A. Sugimoto, K. Shohara, S. Yamanaka, and A.
M. Gabovich, (2010) Tunneling spectroscopy of layered
superconductors: intercalated Li$_{0.48}$(C$_4$H$_8$O)$_x$HfNCl and
De-intercalated HfNCl$_{0.7}$. Euro. Phys. J. B \textbf{73}, 471.

\bibitem{Hiraishi}
M. Hiraishi, R. Kadono, M. Miyazaki, S. Takeshita, Y. Taguchi, Y.
Kasahara, T. Takano, T. Kishiume, and Y. Iwasa, (2010) Anisotropic
superconducting order parameter in Li-intercalated layered
superconductor Li$_x$ZrNCl. Phys. Rev. B \textbf{81}, 014525.

\bibitem{Tou2} 
H. Tou, M. Sera, Y. Maniwa, and S. Yamanaka, (2007) NMR studies of
layered nitride superconductors. Int. J. Mod. Phys. B \textbf{21},
3340.

\bibitem{Tou3}
H. Tou, Y. Maniwa, T. Koiwasaki, and S. Yamanaka, (2001)
Unconventional Superconductivity in Electron-Doped Layered
Li$_{0.48}$(THF)$_y$HfNCl. Phys. Rev. Lett. \textbf{86}, 5775.

\bibitem{Weht} 
R. Weht, A. Filippetti, and W. E. Pickett, (1999) Electron Doping in
the Honeycomb Bilayer Superconductors (Zr, Hf)NCl. Europhys. Lett.
\textbf{48}, 320.

\bibitem{Kitora} 
A. Kitora, Y. Taguchi, and Y. Iwasa, (2007) Probing Electron-Phonon
Interaction in Li$_x$ ZrNCl Superconductors by Raman Scattering. J.
Phys. Soc. Jpn. \textbf{76}, 023706.

\bibitem{Tou4} 
H. Tou, Y. Maniwa, and S. Yamanaka, Superconducting Characteristics
in Electron-Doped Layered Hafnium Nitride: $^{15}$N Isotope Effect
Studies. Phys. Rev. B \textbf{67}, 100509(R) (2003).

\bibitem{Taguchi2} 
Y. Taguchi, T. Kawabata, T. Takano, A. Kitora, K. Kato, M. Takata,
and Y. Iwasa, (2007) Isotope Effect in Li$_x$ZrNCl Superconductors.
Phys. Rev. B \textbf{76}, 064508.

\bibitem{Dompablo}
M. E. Arroyo y de Dompablo, E. Moran, M. A. Alario-Franco, F.
Drymiotis, A. D. Bianchi, and Z. Fisk, (2000) Novel Superconductors
Obtained by Electrochemical Zn Intercalation of $\beta$-ZrNCl and
Related Compounds. Int. J. Inorg. Mater. \textbf{2}, 581.

\bibitem{Zhu2}
L. P. Zhu and S. Yamanaka, (2003) Preparation and Superconductivity
of Chlorine-Deintercalated Crystals $\beta$-MNCl$_{1-x}$ (M = Zr,
Hf). Chem. Mater., \textbf{15}, 1897.

\bibitem{Yamanaka3}
Shoji Yamanaka, (2000) High-$T_{\rm c}$ superconductivity in
electron-doped layered structure nitrides. Annu. Rev. Mater. Sci.
\textbf{30}, 53-82.
\bibitem{Werthamer}
N. R. Werthamer, E. Helfand, and P. C. Hohenberg, (1966) Temperature
and Purity Dependence of the Superconducting Critical Field, Hc2.
III. Electron Spin and Spin-Orbit Effects. Phys. Rev. {\bf 147},
295.


\end{thebibliography}
\end{document}